\documentclass[oneside, a4paper, onecolumn, 11pt]{article}
\usepackage[left=2.5cm,top=2.5cm,bottom=2.5cm,right=2.5cm]{geometry}
\usepackage{lineno}
\usepackage{amsmath}
\usepackage{graphicx}
\usepackage{fancyhdr}
\usepackage{calc}
\usepackage{amssymb}
\usepackage{lipsum}
\usepackage{titlesec}
\usepackage{color}
\usepackage{booktabs}
\usepackage[super,sort&compress,comma]{natbib}
\usepackage{hyperref}
\usepackage{url}
\hypersetup{
	colorlinks=true,
	linkcolor=black,
	filecolor=gray,      
	urlcolor=blue,
	citecolor=black,
}
\usepackage{setspace}
\usepackage{amsfonts}
\usepackage{commath}
\usepackage[innercaption]{sidecap}
\usepackage[framemethod=TikZ]{mdframed}
\usepackage{hyperref}
\usepackage{parskip}
\usepackage{pifont}
\usepackage[labelfont=bf]{caption}
\newcommand{\Rom}[1]{\expandafter\@slowromancap\romannumeral #1@}
\makeatother
{ \normalfont}

\newtheorem{assumption}{Assumption}

\expandafter\def\expandafter\normalsize\expandafter{%
	\normalsize
	\setlength\abovedisplayskip{0pt}
	\setlength\belowdisplayskip{5pt}
	\setlength\abovedisplayshortskip{0pt}
	\setlength\belowdisplayshortskip{5pt}
}

\definecolor{Gray}{gray}{0.75}
\definecolor{changecolour}{rgb}{0, 0, 0.8}

\newmdenv[backgroundcolor=Gray, leftmargin = 0pt, rightmargin = 0pt, linewidth = 0pt, roundcorner = 2 pt, innerleftmargin=5pt, innerrightmargin=5pt, innertopmargin=5pt, innerbottommargin=5pt]{Frame}

\let\oldequation\equation
\let\oldendequation\endequation
\renewenvironment{equation}{\linenomathNonumbers\oldequation}{\oldendequation\endlinenomath}

\let\oldalign\align
\let\oldendalign\endalign
\renewenvironment{align}{\linenomathNonumbers\oldalign}{\oldendalign\endlinenomath}

\let\oldgather\gather
\let\oldendgather\endgather
\renewenvironment{gather}{\linenomathNonumbers\oldgather}{\oldendgather\endlinenomath}

\begin{document}
	
	\newcommand{\kk}{\langle k \rangle}
	\newcommand{\kkk}{\langle k^2 \rangle}
	\newcommand{\er}{Erd\H{o}s-R\'{e}nyi}
	\newcommand{\red}{\color{red}\footnotesize}
	\newcommand{\blue}[1]{{\color{blue} #1}}
	\newcommand{\subfigimg}[3][,]{%
		
		\setbox1=\hbox{\includegraphics[#1]{#3}}
		\leavevmode\rlap{\usebox1}
		\rlap{\hspace*{30pt}\raisebox{\dimexpr\ht1-2\baselineskip}{#2}}
		\phantom{\usebox1}
	}
	
	\linespread{1.4}
	
	\begin{center}
		
		{\LARGE \textbf{Evolutionary dynamics of pairwise and group cooperation in heterogeneous social networks}}
		
		\vspace{2mm}
		Dini Wang$^{1,2,3,4*}$, Peng Yi$^{1,2*}$, Gang Yan$^{2,5}$ and Feng Fu$^{3,6*}$
	\end{center}
	
	\small{
		\begin{enumerate}
			\item
			\textit{College of Electronic and Information Engineering, Tongji University, Shanghai, 201804, China}
			\item
			\textit{Shanghai Research Institute for Intelligent Autonomous Systems, National Key Laboratory of Autonomous Intelligent Unmanned Systems, MOE Frontiers Science Center for Intelligent Autonomous Systems, and Shanghai Key Laboratory of Intelligent Autonomous Systems, Tongji University, Shanghai, 201210, China}
			\item
			\textit{Department of Mathematics, Dartmouth College, Hanover, NH 03755, USA}
			\item
			\textit{Program of Quantitative Social Science, Dartmouth College, Hanover, NH 03755, USA}
			\item
			\textit{School of Physical Science and Engineering, Tongji University, Shanghai, 200092, China}
			\item
			\textit{Department of Biomedical Data Science, Geisel School of Medicine at Dartmouth, Lebanon, NH 03756, USA}
		\end{enumerate}
	}

	\vspace{5mm}

\noindent
\textbf{Understanding how cooperation evolves in structured populations remains a fundamental question across diverse disciplines. 
	The problem of cooperation typically involves pairwise or group interactions among individuals.
	While prior studies have extensively investigated the role of networks in shaping cooperative dynamics, the influence of tie or connection strengths between individuals has not been fully understood.
	Here, we introduce a quenched mean-field based framework for analyzing both pairwise and group dilemmas on any weighted network, providing interpretable conditions required for favoring cooperation.
	Our theoretical advances further motivate us to find that the degree-inverse weighted social ties -- reinforcing tie strengths between peripheral nodes while weakening those between hubs -- robustly promote cooperation in both pairwise and group dilemmas.
	Importantly, this configuration enables heterogeneous networks to outperform homogeneous ones in fixation of cooperation, thereby adding to the conventional view that degree heterogeneity inhibits cooperative behavior under the local stochastic strategy update.
	We further test the generality of degree-inverse weighted social ties in promoting cooperation on $30, 000$ random networks and $13$ empirical networks drawn from real-world systems.
	Finally, we unveil the underlying mechanism by examining the formation and evolution of cooperative ties under social ties with degree-inverse weights.
	Our systematic analyses provide new insights into how the network adjustment of tie strengths can effectively steer structured populations toward cooperative outcomes in biological and social systems.
}

\vspace{5mm}

	\section{Introduction}
	Cooperation is a cornerstone of both biological and societal dynamics~\cite{robert1981, Binmore1991, Levin2020}, recognized as the third fundamental principle of evolution alongside mutation and natural selection~\cite{nowak2011}.
	Cooperative or altruistic behaviors, which contribute to collective welfare at the expense of the individual, are ubiquitous in non-human and human societies, ranging from communal nesting~\cite{riehl2011}, efforts on open-source projects~\cite{chen2022} to engagement in public services~\cite{zhang2022}.
	The rationale behind the emergence of cooperation in a self-interest population has thus captivated substantial interest in diverse disciplines including biology, sociology, and engineering~\cite{jia2019, tang2015, nowak2004, yi2022survey}.
	To study such complex dynamics, evolutionary game theory offers a foundational framework with the donation game and the public goods game as prominent examples of pairwise and group dilemmas respectively~\cite{li2016, nowak2012Evolving, abbass2016,sun2023}.
	However, the unstructured (well-mixed) population generally hinders the proliferation of cooperators unless supported by additional mechanisms~\cite{nowak2004-emergence, nowak2006, ren2024, wei2024}.
	Among others, network reciprocity has received considerable attention, demonstrating that some spatial or structured populations can enhance cooperation by enabling cooperators to cluster and shield themselves from exploitation by defectors~\cite{nowak1992, cheng2015, chiong2012}.
	This naturally leads to a key question: Which structural features of a network most effectively promote the evolution of cooperation?

	In the study of evolutionary games on networks, various evolutionary dynamic rules can lead a structured population to distinct outcomes of strategy assortment.
	Previous research has introduced a number of strategy update mechanisms to describe behavioral evolution, often inspired by patterns observed in real-world interactions~\cite{ramazi2021, wang2024}.
	For instance, locally stochastic update rules~\cite{alex2020}, including death-birth, imitation, and pair-comparison updating, have been extensively explored through computer simulations, analytical calculations~\cite{allen2017} and behavioral experiments~\cite{Arne2010}.
	These update rules drive the evolution of game strategies by allowing individuals to adjust their behavior based on local interactions and influences.
	Under locally stochastic updates, the critical threshold of cooperation on the homogeneous network, where all nodes possess an identical number of neighbors (known as degree), is basically proportional to its node degree~\cite{ohtsuki2006, allen2014}.
	In other words, denser network connections have an adverse effect on the emergence and sustainment of cooperation in terms of homogeneous networks.
	Furthermore, when the average degree is held constant, homogeneous networks hold lower critical thresholds required for supporting cooperation compared to heterogeneous networks, in which node degrees vary~\cite{allen2017, fotouhi2019}.
	This indicates that the high degree heterogeneity of network connections is detrimental to the evolution of cooperation, particularly under locally stochastic updates.

	In real-world systems, individual behavior is shaped not only by the number and distribution of its neighbors but also by the frequency or intensity of their interactions, which are commonly captured by the strengths of social ties~\cite{pajevic2012, basu2016, chen2023}. 
	These differential ties reflect the varying levels of influence, trust, or familiarity between individuals. 
	For instance, in online social media, users interact more frequently with close friends than with those distant acquaintances; in professional settings, communication intensity often varies based on organizational hierarchy or collaborative history~\cite{JP2007}. 
	Accordingly, in the pairwise dilemma, scholars have established a mathematically rigorous condition for the emergence of cooperation on any weighted network and further suggested that strong pairwise ties can facilitate the evolution of collective cooperation through some specific examples~\cite{allen2017, allen2019}.
	The resulting analytical formulation, however, often relies on high-dimensional matrix computations, which not only hinders the application to large-scale networks but also bars the understanding of the interplay between network properties and game parameters.
	The complexity increases further in the group dilemma due to the underlying higher-order (multi-player) interactions, where theoretical analyses are typically limited to either unweighted heterogeneous networks~\cite{alex2020} or weighted but highly symmetric (\textit{e.g.}, node-transitive) network topologies~\cite{Su2019}.
	To uncover general principles governing how weighted social ties alter evolutionary outcomes, there remains a pressing need for the unified and interpretable formulation required for cooperation on any weighted network, capable of balancing expressive simplicity and predictive accuracy.

	Motivated by above considerations, we develop a quenched mean-field framework that yields interpretable thresholds of cooperation in both pairwise and group dilemmas, applicable for arbitrary networks with heterogeneous social ties.
	Subsequently, we validate the accuracy of analytical results by Monte Carlo simulations on a variety of synthetic networks, and specifically by comparisons with the exact solution from Allen, et al~\cite{allen2017} across networks of varying size and topology.
	We demonstrate that degree-inverse weighted social ties, where the strength of an edge is proportional to the inverse product of the unweighted degrees of its endpoints, can substantially lower the thresholds of cooperation in both pairwise and group dilemmas.
	We then surprisingly find that degree-inverse weighted social ties can not only facilitate the emergence of cooperation, but also extend the conventional understanding that high degree heterogeneity inhibits cooperative behavior.
	To assess the generality of this finding, we compare the thresholds of cooperation between homogeneous social ties and degree-inverse weighted ones, and examine $13$ empirical networks drawn from social, Facebook, retweet, contact, email, and collaboration domains.
	Finally, we provide a mechanistic explanation of how degree-inverse weighted social ties promote cooperation from the perspective of the formation and evolution of cooperation ties throughout evolutionary dynamics.
	
	The rest of the manuscript is organized as follows. 
	Section II provides the model formulation, involving the population structure, games and payoffs, as well as strategy evolution dynamics.
	Section III introduces the mathematical framework for the critical thresholds of cooperation in both the donation game and the public goods game on arbitrary weighted networks.
	Section IV demonstrates the accuracy of theoretical analyses ($A$), proposes that degree-inverse weighted social ties can strongly promote pairwise and group cooperation ($B$ and $C$), tests the generality of this setting ($D$), and explain its underlying mechanism from the viewpoint of cooperation ties ($E$).
	Finally, the conclusion and discussion are presented in Section V.

	\section{Model formulation}
	\subsection{Population structure} 
	We mainly focus on the heterogeneous population, where individuals have distinct numbers of neighbors and social ties are of varying strengths (Figure~\ref{fig1}(a)).
	To quantify the intrinsic properties of such a structured population, we introduce a connected, undirected and weighted network, $\mathcal{G}(\mathcal{N}, \mathcal{E})$, with $\mathcal{N}$ representing a set $N$ nodes and $\mathcal{E}$ indicating a set of unordered pairs of these nodes called edges.
	In an edge, the nodes at both ends should be distinct, as we do not consider the self-loop from one node to itself.
	Here each node represents an individual of the population and every edge implies a social tie whose interaction frequency can be measured by the edge weight $w_{ij}$ for $i, j \in \mathcal{N}$.
	Notably, due to the undirected setting of edges, $w_{ij} = w_{ji}$.

	\begin{assumption}
		The network $\mathcal{G}$ is connected.
		\label{ass1}
	\end{assumption}
	
	Assumption~\ref{ass1} implies that every node in $\mathcal{G}$ can create a lineage that takes over the whole population in evolutionary dynamics.
	
	Regarding the weighted network $\mathcal{G}$, there exist two fundamental topological properties: one is the (unweighted) degree of a node $i \in \mathcal{N}$, denoted as $d_i$, equal to the number of the node $i$'s neighbors; the other is the weighted degree of a node $i$ that accumulates the weights of edges connected to $i$, defined as $k_i  = \sum_{j \in \mathcal{N}} w_{ij}$.
	Besides these, we introduce the random walk on the network -- a walker located on the node $i$ of $\mathcal{G}$ moves to the immediate neighbor $j$ with the probability proportional to the edge weight.
	To formulate this process, the transition probability from node $i$ to node $j$ on the network can be defined as $p_{ij} = w_{ij} / k_i$, and all the possible combinations of $(i,j)$ constitute the transition matrix of the random walk, denoted as $\mathbf{P}=(p_{ij})_{i, j \in \mathcal{N}}$.
	Moreover, the $n$-step probability from $i$ to $j$ on $\mathcal{G}$ is represented by $p_{ij}^{(n)} = [\mathbf{P}^n]_{ij}$.
	When the step $n$ is sufficiently large, there exists a stationary distribution $\{ \pi_i \}_{i \in \mathcal{N}}$ for the random walk on the network, in which $\pi_i = k_i / \sum_{i \in \mathcal{N}} k_i$ equals the normalized weighted degree of the node $i$~\cite{allen2017}.
	Generally speaking, all the transition probabilities with sufficient steps terminating at the node $j$ equal $\pi_j$, or mathematically $\lim\limits_ { n \rightarrow \infty } p_{ij} = \pi_j$ for any $i, j \in \mathcal{N}$.
	
	To simplify the mathematical expression, we define the shorthand notation for any function $g_i$ on the network $\mathcal{G}$:
	$$
	g_i^{(n)} = \sum_{j \in \mathcal{N}} p_{ij}^{(n)} g_j
	$$
	which means the expected value averaged by the $n$-step neighbors of the focal individual $i$.

	\subsection{Games and payoffs} 
\begin{figure*}[!t]
	\centering
	\includegraphics[width=\textwidth]{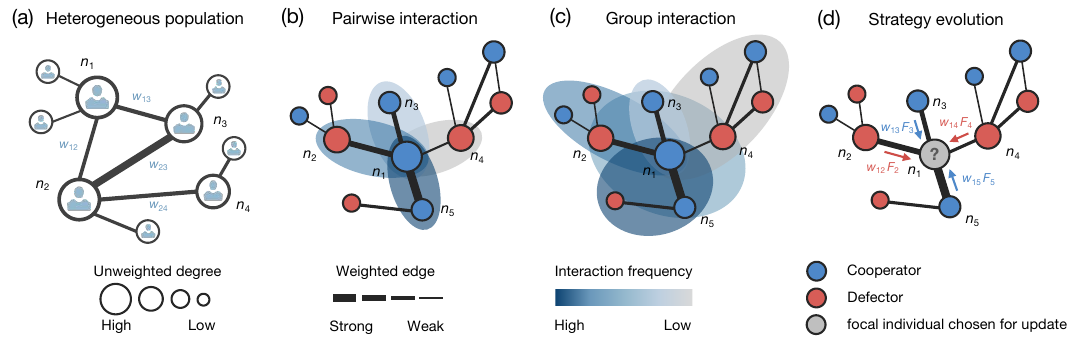}
	\caption{Evolutionary dynamics of cooperation based on pairwise and group interactions in a heterogeneous population.
		(a), A heterogeneous population is modeled as a weighted network, where nodes have different numbers of neighbors, defined as (unweighted) degrees, and edges between pairs of connected nodes have varying weights, denoted as $w_{ij}$ for nodes $i$ and $j$.
		Specifically, the size of a node reflects its unweighted degree and the width of an edge exhibits its strength.
		(b), The pairwise interaction on the network occurs on the edge between two connected nodes (cooperators colored in blue and defectors in red).
		The color intensity shows the interaction frequency that depends on the edge weight.
		(c), The group interaction on the network involves a focal node and all of its neighbors with the edge weight measuring the corresponding interaction frequency.
		(d), The gray node $n_1$ is selected to update its strategy, and subsequently one of its neighbor $n_i$ competes for this vacancy with the probability proportional to the fitness $F_i$ and the edge weight $w_{1i}$.}
	\label{fig1}
\end{figure*}

	In reality, social dilemmas are generally manifested in the patterns of either pairwise or group interactions.
	Specifically, in a network-structured population, a pairwise interaction  merely involves the focal individual (Figure~\ref{fig1}(b)) and one of its network neighbors, while a group interaction encompasses the focal individual and all of its immediate neighbors (Figure~\ref{fig1}(c)).
	
	Here, we adopt the donation game and the public goods game to model the pairwise and group dilemmas respectively.
	There are two strategies: cooperation (C) and defection (D).
	For the network game, the individual state of a population can be delineated as a binary vector $\mathbf{s} = (s_i)_{i \in \mathcal{N}} \in \left\{0,1\right\}^{\mathcal{N}}$, where $1$ corresponds to cooperation and $0$ to defection.

\subsubsection{Donation game}
In a round of the donation game, the cooperator pays a cost $c$ and thus creates a benefit $b$ for its recipient, while the defector makes no contributions.
Even though unilateral defection can protect one's benefit, mutual cooperation optimizes the bilateral welfare, therefore inducing the cooperation dilemma~\cite{nowak2006, nowak2012Evolving}.

Regarding the network game scenario, the individual $i$ engages in the pairwise game with its direct neighbor $j$ and the interaction frequency $w_{ij}$, thus acquiring the payoff
\begin{equation}
	f_{i(j)}(\mathbf{s})= bw_{ij}s_{j}  - cw_{ij}s_{i} .
\end{equation}
Subsequently, the payoffs from frequency-heterogeneous interactions with all the accessible neighbors should be normalized by the weighted degree, and the individual $i$'s average payoff is accordingly calculated as
\begin{equation}
	f_{i} (\textbf{s})
	= \frac{\sum_{j \in \mathcal{N}} f_{i(j)}(\textbf{s})}{k_{i}}
	= b\sum_{j \in \mathcal{N}} p_{ij} s_{j} - cs_{i}.
\end{equation}

\subsubsection{Public goods game}
Beyond the dyadic interaction, the public goods game involves a group of players, where each cooperator invests a cost $c$ to the public pool and thereafter the total investment is multiplied by the enhancement factor $r$.
The collective benefit generated is then shared by all the players.

Considering the networked public goods game, each individual hosts a game including itself and all of its direct neighbors.
The interaction frequency between them, equal to the corresponding edge weight, measures the amount of investment and allocation~\cite{Su2019}, while that of the focal individual relies on the average weights of edges connected to it.
The conventional unweighted setting can be recovered by assuming the identical frequency for each interaction~\cite{santos2008}.

For analysis of public goods games on networks, we construct a network $\hat{\mathcal{G}}$ by simply adding a self-loop to each node $i$ of the original $\mathcal{G}$.
Therefore, the weight of the edge between two distinct nodes ($i, j \in \hat{\mathcal{G}}$ and $i \neq j$) maintains $\hat{w}_{ij} = {w}_{ij}$, while the weight of each self-loop is defined as $\hat{w}_{ii} = k_i / d_i$, which equals the average weight of the edges between the node $i$ and its neighbors.
Furthermore, the unweighted degree and the weighted degree of a node $i \in \mathcal{G}$ can be represented by $\hat{d}_i = d_i + 1$ and $\hat{k}_i = k_i + \hat{w}_{ii}$, respectively.
In particular, if $w_{ij} = 1$ (\textit{i.e.}, the unweighted network), then $\hat{k}_i = \hat{d}_i$ corresponds to the conventional setting~\cite{santos2008} and $\hat{w}_{ii} = 1$ means a self-interaction in the game hosted by $i$.
For the random walk on $\hat{\mathcal{G}}$, the transition matrix is represented by $\hat{\mathbf{P}}=(\hat{p}_{ij})_{i, j \in \mathcal{N}}$, where $\hat{p}_{ij} = \hat{w}_{ij}/ \hat{k}_i$.
Its $n$-step version is $\hat{\mathbf{P}}^n$ with the $(i, j)$-element $\hat{p}_{ij}^{(n)}$.
More generally, we envision a pattern of $n$-step random walks on $\mathcal{G}$ followed by $m$-step random walks on $\hat{\mathcal{G}}$, and the transition probability from $i$ to $j$ is denoted as $p^{(n,m)}_{ij} = \sum_{y \in \mathcal{N}} p_{iy}^{(n)} \hat{p}_{yj}^{(m)}$.

On the network $\hat{\mathcal{G}}$, the individual $i$ joins in the game hosted by its neighbor $j$ ($j=i$ is allowed) with the interaction frequency $\hat{w}_{ij}$, and thus gets the payoff as
\begin{equation}
	\begin{aligned}
		f_{i(j)}(\mathbf{s}) 
		=& r \frac{\hat{w}_{ij}}{\hat{k}_j} \sum_{y \in {\mathcal{N}}} \hat{w}_{jy} s_y - \hat{w}_{ij} s_i \\
		=& r \hat{w}_{ij} \sum_{y \in {\mathcal{N}}} \hat{p}_{jy} s_y - \hat{w}_{ij} s_i.
	\end{aligned}
\end{equation}
Then all the payoffs of the games the individual $i$ participates in are normalized by its weighted degree $\hat{k}_i$:
\begin{equation}
	f_{i} (\textbf{s})
	= \frac{\sum_{j \in {\mathcal{N}}} f_{i(j)}(\textbf{s})}{\hat{k}_{i}}
	= r \sum_{y \in {\mathcal{N}}} \hat{p}_{iy}^{(2)} s_{y} -s_{i}.
	\label{payoff-pgg}
\end{equation}

The fitness of individual $i$ in both the networked donation game and public goods game is given by $F_{i}(\mathbf{s})=1+\delta f_{i}(\mathbf{s})$, where $\delta  > 0$ quantifies the strength of selection~\cite{wu2010}. In this work, we focus on the weak selection regime, indicating that $\delta \rightarrow 0$, but simulations go beyond this limit.

\subsection{Strategy evolution dynamics}
We consider the strategy evolution mechanism driving the behavioral dynamics as the death-birth update, where a to-be-updated individual is randomly selected and then imitates one of its neighbor's behavior by evaluating the fitness weighted by the corresponding interaction frequency (Figure~\ref{fig1}(d)).
Thus, under the death-birth mechanism~\cite{alex2020}, the probability that the individual $i$ copies its networked neighbor $j$'s strategy to update oneself can be quantified as
\begin{equation}
	\mathrm{Pr} \left[j\rightarrow i\right]\left(\mathbf{s}\right) 
	= \frac{ {w}_{ij}F_{j} (\mathbf{s})}{\sum_{y \in \mathcal{N}} {w}_{iy}F_{y}(\mathbf{s}) } .
	\label{trans-death-birth}
\end{equation}
where $\sum_{j \in \mathcal{N}} \mathrm{Pr} \left[j\rightarrow i\right]\left(\mathbf{s}\right) = 1$ ensures the overall probability that the individual $i$ is replaced equals 1.

\begin{assumption}
	For any configuration of $\mathbf{s}$, the transition probability (\ref{trans-death-birth}) is a smooth function with respect to the selection strength $\delta$ in a small neighborhood of $\delta = 0$~\cite{tarnita2009}.
	\label{ass2}
\end{assumption}

\section{Mathematical framework}
We establish a mathematical framework for interpretable thresholds of cooperation in both pairwise and group dilemmas, applicable for arbitrary networks having social ties with heterogeneous weights.
In this framework, we firstly implement the weak-selection perturbation of fixation probability to split the neutral drift (\textit{i.e.}, $\delta = 0$) out of the natural selection.
As for the neutral drift, we adopt the quenched mean-field approximation, which assumes the absence of dynamical correlations between neighboring strategies. 
We further integrate this approximation within the coalescing random walk that is a classical treatment in evolutionary theory used to trace ancestral lineages.
Through some mathematical transformations, we obtain the approximate closed-form conditions of both pairwise and group cooperation for arbitrary network structures.

\subsection{Weak-selection perturbation of fixation probability}
Our central aim is identifying if a specific network can support cooperation to evolve within pairwise and group interactions; and if so, by how much.
When biological mutation is not taken into consideration, the system will eventually reach absorption, either all-cooperator or all-defector, through sufficient rounds of evolution~\cite{alex2020, allen2017}.
The fixation probability of cooperation, $\rho_{\mathrm{C}}$, describes the potential that a single cooperator takes over the entire population when the system is fixed.
On neutral drift (\textit{i.e.}, $\delta = 0$), this fixation probability is equal to $1/N$~\cite{tarnita2009}.
Therefore, the condition that $\rho_{\mathrm{C}} > 1/N$ determines the dominant role of cooperators compared with defectors.

To formulate the dynamical evolution, we employ a continuous-time Markov chain $\mathbf{s}(t) = (s_1(t), s_2(t), ..., s_N(t))'$, where $t \geq 0$.
Here we start with the initial state as a cooperator occupying the node $u$ of a networked population of defectors, meaning that $s_u^u(0) = 1$ while $s_i^i(0) = 0$ for $i \in \mathcal{G}$ and $ i \neq u$.
Under such an initial configuration $u$, the fixation probability of cooperation is equal to its expected reproductive-value-weighted frequency in the limit of $t \rightarrow \infty$~\cite{allen2017}, thus yielding
\begin{equation}
	\rho^{u}_{\mathrm{C}} := \lim\limits_{t \rightarrow  \infty  }  \mathbb{E}  \Bigg[ \sum_{i \in \mathcal{N}}\pi_i s_i^{u}(t) \Bigg],
	\label{fix_org}
\end{equation}
where the superscript $u$ represents the initial configuration $u$.

Subsequently, we apply the fundamental theorem of calculus to the fixation probability (\ref{fix_org}) to produce
\begin{equation}
	\rho^{u}_{\mathrm{C}} =  \pi_u + \int_{0}^{\infty} \mathbb{E} \Bigg[  \sum_{i \in \mathcal{N}}\pi_i \cdot \dfrac{ds_i^{u}(t)}{dt} \Bigg]  dt.
\end{equation}

As such, averaging the fixation probability for all the possible initial configurations of a single mutant of the cooperator across the population can lead to that for the initial configuration of a random mutant, denoted as
\begin{equation}
	\rho_{\mathrm{C}} =  \frac{1}{N} + \int_{0}^{\infty} \mathbb{E} \Bigg[  \sum_{i,u \in \mathcal{N}}\pi_i \cdot \dfrac{ds_i^{u}(t)}{dt} \Bigg]  dt.
	\label{rho-def}
\end{equation}

For a certain state $\mathbf{s}$, the reproductive-value-weighted transition change of the individual state in one step assuming proper scaling in the continuum limit can be derived as
\begin{equation}
	\begin{aligned}
		\sum_{i \in \mathcal{N}} \pi_i \cdot \dfrac{ds_{i}}{dt} 
		= & \sum_{i \in \mathcal{N}} \pi_i \biggl(\sum_{j \in \mathcal{N}}   \text{Pr}[j \rightarrow i] \left(\mathbf{s} \right)\cdot s_j - s_i \biggr)   \\
		= & \delta \cdot  \sum_{i \in \mathcal{N}}   \pi_i s_i  \biggl(  f_i(\mathbf{s})  - f_i^{(2)}(\mathbf{s}) \biggr)  + \mathcal{O}(\delta^2)  \\
	\end{aligned},
	\label{change}
\end{equation}
where $f_i^{(2)}(\mathbf{s})$ represents the expected fitness of two-step neighbors of the node $i$ under the state $\mathbf{s}$.
Here, Eq.~(\ref{change}) is achieved by the Taylor expansion regarding $\delta = 0$ based on the differentiability from assumption~\ref{ass2}. 
For the detailed derivation, please refer to Appendix \textit{A}.

By substituting the individual payoff of the donation game into the one-step transition change (\ref{change}), the fixation probability of pairwise cooperation can be recast into
\begin{equation}
	\begin{aligned}
		\rho_{\mathrm{C}}  = & \frac{1}{N} + \delta  \int _ { 0 } ^ {  \infty} \mathbb{E}  \left[ - c \biggl(\sum_{i,u \in  \mathcal{N}}\pi_{i} s_{i}^{u}(t)  s_{i}^{u}(t) -\sum_{i,u \in  \mathcal{N}}\pi_{i} s_{i}^{u}(t) s_{i}^{u}(t)^{(2)} \biggr)  \right. \\
		& \left.  +  b \biggl( \sum_{i,u \in \mathcal{N}}\pi_{i} s_{i}^{u}(t)  s_{i}^{u}(t)^{(1)} - \sum_{i,u \in \mathcal{N}}\pi_{i} s_{i}^{u}(t) s_{i}^{u}(t)^{(3)} \biggr) \right]_{\delta=0} dt + \mathcal{O}(\delta ^{2}).
	\end{aligned}\label{perturb-1}
\end{equation}
where $s_{i}^{u}(t)^{(n)}$ indicates the expected state of $n$-step neighbors of the node $i$ at time $t$ under the initial configuration $u$.

Similarly, regarding the public goods game, the fixation probability of group cooperation is
\begin{equation}
	\begin{aligned}
		\rho_{\mathrm{C}}  = & \frac{1}{N} + \delta  \int _ { 0 } ^ {  \infty} \mathbb{E}  \left[ -  \biggl(\sum_{i,u \in  \mathcal{N}}\pi_{i} s_{i}^{u}(t)  s_{i}^{u}(t) -\sum_{i,u \in  \mathcal{N}}\pi_{i} s_{i}^{u}(t) s_{i}^{u}(t)^{(2,0)} \biggr)  \right. \\
		& \left.  +  r \biggl( \sum_{i,u \in \mathcal{N}}\pi_{i} s_{i}^{u}(t)  s_{i}^{u}(t)^{(0,2)} - \sum_{i,u \in \mathcal{N}}\pi_{i} s_{i}^{u}(t) s_{i}^{u}(t)^{(2,2)} \biggr) \right]_{\delta=0} dt + \mathcal{O}(\delta ^{2}).
	\end{aligned}\label{perturb-2}
\end{equation}

\subsection{Quenched mean-field approximation}
To untangle the joint states in the fixation probabilities~(\ref{perturb-1}) and (\ref{perturb-2}), we introduce the quenched mean-field approximation, which is employed by neglecting the dynamical correlations between the states of the neighbors.
To this end, we firstly perform the state transformation in the probabilistic sense by introducing a probability vector $\mathbf{x}^{u}(t)=(x_1^{u}(t), x_2^{u}(t), \ldots, x_N^{u}(t))'$ for $t \geq 0$, where $x_i^{u}(t)$ represents the probability of the individual $i$ to cooperate given the initial configuration $u$ at time $t$. 
Here $0 \le x_i^{u}(t) \le 1$, and $x_i^{u}(t)=1$ corresponds to a cooperator while $x_i^{u}(t)=0$ to a defector. 
For a round of the strategy update, the state of the population changes according to the below formula \cite{asava2001, tan2015}:
\begin{equation}
	\begin{aligned}
		&\mathbf{x}^{u}(t+1) = \mathbf{Pr}(\mathbf{s}^{u}(t)) \cdot \mathbf{s}^{u}(t) \\
		&\mathbf{s}^{u}(t+1) = \mathcal{R}(\mathbf{x}^{u}(t+1))
	\end{aligned},
	\label{prob model}
\end{equation}
where $\mathbf{Pr}(\mathbf{s})$ is the matrix of the state transition given the state $\mathbf{s}$ with the $(i,j)$-element $\mathbf{Pr}(\mathbf{s})_{ij} = \mathrm{Pr} \left[j\rightarrow i\right] (\mathbf{s})$; the operator $\mathcal{R}(\cdot)$ is a realization of the probability vector in the bracket. 
Specifically, $s_i$ is set as 1 with the probability of $x_i$ or 0 with the probability of $1-x_i$ for $i \in \mathcal{G}$.
Hence, $\mathbf{x}^{u}(t) = \mathbb{E}(\mathbf{s}^{u}(t))$ across the timescale.

Subsequently, we move forward to the individual-based quenched mean-field approximation for the neutral drift, that is
\begin{equation}
	\mathbb{E} \left[ s_{i}^{u}(t) s_{j}^{u}(t) \right]_{\delta=0} \approx \left[x_{i}^{u}(t)\right]_{\delta=0} \cdot  \left[x_{j}^{u}(t)\right]_{\delta=0}.
	\label{mean}
\end{equation}
Then substituting the approximation (\ref{mean}) into the weak-selection perturbative expansion of fixation probability (\ref{perturb-1}) therefore produces the condition for pairwise cooperation (\textit{i.e.}, $\rho_{\mathrm{C}}> 1/N $) as
\begin{equation}
	\begin{aligned}
		& b \int _ { 0 } ^ {  \infty}   \biggl( \sum_{i,u \in  \mathcal{N}}\pi_{i} x_{i}^{u}(t)  x_{i}^{u}(t)^{(1)} - \sum_{i,u \in  \mathcal{N}}\pi_{i} x_{i}^{u}(t) x_{i}^{u}(t)^{(3)} \biggr)  _{\delta=0} dt \\
		& > c \int _ { 0 } ^ {  \infty}       \biggl(\sum_{i,u \in  \mathcal{N}}\pi_{i} x_{i}^{u}(t)  x_{i}^{u}(t) -\sum_{i,u \in  \mathcal{N}}\pi_{i} x_{i}^{u}(t) x_{i}^{u}(t)^{(2)} \biggr)   _{\delta=0} dt
		\label{condition-pair-1}
	\end{aligned},
\end{equation}
over which pairwise cooperation is favored to replace defection across the population.

Analogously, the condition for group cooperation can be transformed into
\begin{equation}
	\begin{aligned}
		& r \int _ { 0 } ^ {  \infty}   \biggl( \sum_{i,u \in  \mathcal{N}}\pi_{i} x_{i}^{u}(t)  x_{i}^{u}(t)^{(0,2)} - \sum_{i,u \in  \mathcal{N}}\pi_{i} x_{i}^{u}(t) x_{i}^{u}(t)^{(2,2)} \biggr)  _{\delta=0} dt \\
		& > \int _ { 0 } ^ {  \infty}       \biggl(\sum_{i,u \in  \mathcal{N}}\pi_{i} x_{i}^{u}(t)  x_{i}^{u}(t) -\sum_{i,u \in  \mathcal{N}}\pi_{i} x_{i}^{u}(t) x_{i}^{u}(t)^{(2,0)} \biggr)   _{\delta=0} dt
		\label{condition-group-1}
	\end{aligned}.
\end{equation}

\subsection{Coalescing random walks}
In what follows, we focus on the neutral drift, where the game fitness do not vary among individuals, to unpack the probabilistic states in the condition for cooperation~(\ref{condition-pair-1}).
As such, we apply the coalescing random walk to trace the ancestor along the lineage backwards to the initial state.
Regarding the neutral drift ($\delta = 0$), the probability that the individual $i$ copies $j$'s state driven by the death-birth update for any initial configuration $u$ and any time $t$ is
\begin{equation}
	\left[\text{Pr}[j \rightarrow i](\mathbf{x}^u(t)) \right]_{\delta=0} = {p}_{ij}.
	\label{coalesence-eq1}
\end{equation}
In essence, the strategy shift is a reverse process of the networked random walk for each pair of connected nodes $(i, j)$.

Since merely a single individual changes the strategy in each timestep, the one-step transition probability of the state can be delineated as
\begin{equation} 
	{\xi}_{ij} = \begin{cases}1 - \dfrac{1}{N}  & i = j \\ \dfrac{1}{N} {p}_{ij} & i \neq j\end{cases}.
	\label{xi}
\end{equation}
The one-step state transition matrix of the death-birth update is thus equal to
\begin{equation}
	{\Xi} = \frac{1}{N}  {P} + \left(1-\frac{1}{N} \right) E,
	\label{Xi}
\end{equation}
where $E$ is the identity matrix. 
Accordingly, it is straightforward to get $ {P} = N  {\Xi}- \left(N-1 \right)E$, and the stationary distribution $\{ \pi_{i} \}_{i \in \mathcal{N}}$ is applicable for matrices $ {\Xi}$ and ${P}$.
To match the continuum Markov chain, we consider the continuous version of the transition matrix as $\Xi(t) = (\xi_{ij}(t))_{i,j \in \mathcal{N}}$, which also maintains the reversibility property that $\pi_i  {\xi}_{ij}{(t)} = \pi_j  {\xi}_{ji}{(t)}$.

The current state of the population is determined by the transition probability on the neutral drift as well as the initial configuration of the state, quantified as
\begin{equation}
	\left[ \textbf{x}^u(t) \right] _{\delta = 0} =  {\Xi}{(t)}  \cdot \left[  \textbf{x}^u(0)\right] _{\delta = 0} .
\end{equation}
From the individual viewpoint, the individual $i$'s state under the initial configuration $u$ at time $t$ should be dated back towards the initial mutant $u$.
Since $x_u^u(0) = 1$, the individual $i$'s state can be denoted as
\begin{equation}
	\left[x_i^u(t) \right] _{\delta = 0} = \sum_{j \in \mathcal{N}}  {\xi}_{ij}{(t)}  x_j^u(0)  =   {\xi}_{iu}{(t)}   x_u^u(0) =  {\xi}_{iu}{(t)}.
	\label{trace}
\end{equation}

Next, we substitute Eq. (\ref{trace}) into the condition~(\ref{condition-pair-1}) and then make some mathematical transformation to the fundamental piece, thus yielding
\begin{equation}
		  \int _ { 0 } ^ {  \infty} \sum_{i,u \in \mathcal{N}}\left[   \pi_{i} x_{i}^{u}(t) x_{i}^{u}(t)^{(n)} \right]  _{\delta = 0}dt =  \frac{1}{2} \int_{0}^{\infty}  \sum_{i,u \in \mathcal{N}}  \pi_{u}  {\mathcal{\xi}}_{ui}{(t)}  {p}_{iu}^{(n)}   dt,
	\label{approx-right}
\end{equation}
where $n \in \mathbb{N}$ and the detailed derivation is shown in Appendix \textit{B}.

Substitute Eq.~(\ref{approx-right}) into the condition for pairwise cooperation~(\ref{condition-pair-1}), and transform this into the discrete-time version to recast the condition for pairwise cooperation into
\begin{equation}
		b \biggl( \sum_{t = 0}^{\infty} \sum_{i,u \in \mathcal{N}} \pi_u   {\xi}_{ui}^{(t)}  p_{iu}^{(1)}  - \sum_{t = 0}^{\infty} \sum_{i,u \in \mathcal{N}} \pi_u   {\xi}_{ui}^{(t)}  p_{iu}^{(3)}  \biggr) >  c \biggl( \sum_{t = 0}^{\infty} \sum_{i,u \in \mathcal{N}} \pi_u   {\xi}_{uu}^{(t)} - \sum_{t = 0}^{\infty} \sum_{i,u \in \mathcal{N}} \pi_u   {\xi}_{ui}^{(t)}  p_{iu}^{(2)}  \biggr)
		\label{condition-pair-2}
\end{equation}

The group scenario is more complicated than the pairwise one as it needs extra approximations (see Appendix \textit{C} for details).
The condition for group cooperation can be thus reorganized as
\begin{equation}
	r \biggl( \sum_{t = 0}^{\infty} \sum_{i,u \in \mathcal{N}} \pi_u   {\xi}_{ui}^{(t)}  p_{iu}^{(0,2)}  - \sum_{t = 0}^{\infty} \sum_{i,u \in \mathcal{N}} \pi_u   {\xi}_{ui}^{(t)}  p_{iu}^{(2,2)}  \biggr) >  \biggl( \sum_{t = 0}^{\infty} \sum_{i,u \in \mathcal{N}} \pi_u   {\xi}_{uu}^{(t)} - \sum_{t = 0}^{\infty} \sum_{i,u \in \mathcal{N}} \pi_u   {\xi}_{ui}^{(t)}  p_{iu}^{(2,0)}  \biggr)
		\label{condition-group-2}
\end{equation}

\subsection{Critical thresholds for cooperation}
By eliminating the infinite summation, we get the condition for cooperation in the pairwise dilemma is
\begin{equation}
		b \left( \sum_{u \in \mathcal{N}} \pi_u p_{uu}^{(1)} + \sum_{u \in \mathcal{N}} \pi_u p_{uu}^{(2)} - 2 \sum_{u \in \mathcal{N}} \pi_u^2 \right)  > c \left( \sum_{u \in \mathcal{N}} \pi_u p_{uu}^{(0)} + \sum_{u \in \mathcal{N}} \pi_u p_{uu}^{(1)} - 2 \sum_{u \in \mathcal{N}} \pi_u^2 \right).
	\label{condition-pair-3}
\end{equation}
Please refer to Appendix \textit{D} for details of mathematical transformation.

As self-loops are not taken into consideration in $\mathcal{G}$, it is natural to obtain
\begin{equation*}
	\sum_{u \in \mathcal{N}} \pi_u p_{uu}^{(0)} = 1,  \sum_{u \in \mathcal{N}} \pi_u p_{uu}^{(1)}  =0,  \sum_{u \in \mathcal{N}} \pi_u^2  =\frac{\eta_k}{N},
\end{equation*}
where $\eta_k = \langle k^2 \rangle / \langle k \rangle^2$ represents the weighted degree heterogeneity with $\langle k \rangle = \sum_{i \in \mathcal{N}} k_i / N$ denoting the first moment of the weighted degree and $\langle k^2 \rangle = \sum_{i \in \mathcal{N}} k_i^2 / N$ representing the second moment of the weighted degree.

The theoretical threshold of cooperation in the pairwise social dilemma is
\begin{equation}
	B^* = \frac{N - 2 \eta_k}{N + N \sum_{u \in \mathcal{N}} \pi_u p_{uu}^{(2)} - 2 \eta_k}
	\label{bc_star}
\end{equation}
over which pairwise cooperation is favored to spread in the population.

In the similar manner, considering the group dilemma, the condition for group cooperation can be reorganized as
\begin{equation}
		r \left( \sum_{u \in \mathcal{N}} \pi_u p_{uu}^{(0,2)} + \sum_{u \in \mathcal{N}} \pi_u p_{uu}^{(1,2)} - 2 \sum_{u,i \in \mathcal{N}} \pi_u {p}_{ui}^{(0,2)} \pi_i \right)  > \sum_{u \in \mathcal{N}} \pi_u p_{uu}^{(0,0)} + \sum_{u \in \mathcal{N}} \pi_u p_{uu}^{(1,0)} - 2 \sum_{u \in \mathcal{N}} \pi_u^2 .
\end{equation}

The theoretical threshold of cooperation in the group social dilemma is given by
\begin{equation}
	r^* = 
	\frac{N - 2\eta_k}{ N \sum_{u \in \mathcal{N}} \pi_u p_{uu}^{(0,2)} + N \sum_{u \in \mathcal{N}} \pi_u p_{uu}^{(1,2)} - \frac{2d^{(0,2)}}{\langle k \rangle} - 2 \eta_k  }.
	\label{r_star}
\end{equation}
where $d^{(0,2)} = \sum_{u \in \mathcal{N}}(\pi_u - \sum_{i \in \mathcal{N}} p_{ui}^{(0,2)} \pi_i)$ can measure the heterogeneity of the weighted degree distribution.

\section{Main results}
\subsection{Validation of theoretical results}
\begin{figure*}[!t]
	\centering
	\includegraphics[width=0.66\textwidth]{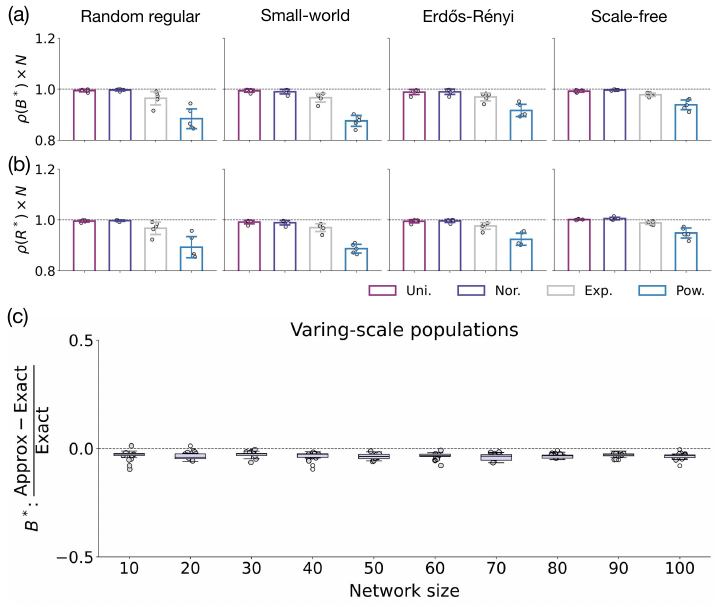}
	\caption{Simulation validation of theoretical critical thresholds of both pairwise and group cooperation.
		(a)-(b), Box plots of size-normalized fixation probabilities from $2 \times 10^6$ replicated simulations under our approximated thresholds of the benefit-to-cost ratio $B^*$ (a) and the enhancement factor $R^*$ (b), derived from Eqs. (\ref{bc_star}) and (\ref{r_star}).
		We apply four types of synthetic networks of size $N = 100$ and average degrees $\langle d \rangle = 4, 8, 10, 12$, each coupled with edge weights drawn from uniform (Uni.), normal (Nor.), exponential (Exp.), and power-law (Pow.) distributions.
		The intensity of selection is adopted as $\delta = 0.025$ in every simulation.
		(c), Relative difference between the approximated $B^*$ obtained from Eq. (\ref{bc_star}) and the exact $B^*$ from Eq. (\ref{exact}) on varying-scale networks of average degree $\langle d \rangle = 4$ with edges weighted by the uniform distribution.}
	\label{fig2}
\end{figure*}

We validate the accuracy of our approximations for the critical thresholds of both pairwise and group cooperation.
Firstly, we replicate $10^6$ Monte Carlo simulations of fixation probabilities by substituting the theoretical thresholds, $B^*$ from Eq.~(\ref{bc_star}) and $R^*$ from Eq.~(\ref{r_star}), into the game parameters, $b/c$ and $r$, respectively, as illustrated in Figure~\ref{fig2}(a) and~\ref{fig2}(b).
If the resulting fixation probabilities normalized by the population size approach $1$, this indicates strong alignment between analytical predictions and simulation outcomes.
To assess the generality of theoretical predictions, we examine four representative types of random networks: random regular (RR), small-world (SW)~\cite{newman1999}, Erd\"os-R\'enyi (ER)~\cite{erdos1959}, and scale-free (SF)~\cite{barabasi1999}, each with a population size of 100, average unweighted degrees ranging from $4$ to $12$, and diverse configurations of edge weights.
As shown in Figure~\ref{fig2}(a) and~\ref{fig2}(b), our theoretical results exhibit high accuracy for both pairwise and group cooperation under uniform and normal distributions of edge weights across all network types.
Meanwhile, under exponential edge weight distributions, the predictions remain fairly accurate, achieving agreement levels exceeding 95\%.
In contrast, when edge weights follow a power-law distribution, the predictions remain reasonably accurate for ER and SF networks but show a slight deviation—approximately 10\%—from simulated results in RR and SW networks.
Overall, these results demonstrate that our approximations of the critical thresholds for pairwise and group cooperation are sufficiently accurate and reliable for use in subsequent analyses.

Secondly, we compare our approximation for the critical threshold of pairwise cooperation, $B^*$, with the exact solution derived from pure coalescence theory, given by
\begin{equation}
	B^* = \frac{\sum_{i,j \in \mathcal{N}} \pi_i p_{ij}^{(2)} \tau_{ij}}{\sum_{i,j \in \mathcal{N}} \pi_i p_{ij}^{(3)} \tau_{ij} - \sum_{i,j \in \mathcal{N}} \pi_i p_{ij} \tau_{ij}},
	\label{exact}
\end{equation}
where the coalescence time $\tau_{ij}$ satisfies the recurrence relation
\begin{equation}
	\tau_{ij} =  
	\begin{cases}
		0 & i = j\\
		1 + \frac{1}{2} \sum_{\iota \in \mathcal{N}} (p_{i \iota} \tau_{j \iota} + p_{j \iota} \tau_{i \iota}) & i \neq j
	\end{cases}.
\end{equation}
To perform this comparison, we employ synthetic networks with uniformly distributed edge weights and varying sizes from 10 to 100.
As shown in Figure~{\ref{fig2}}(c), it globally witnesses a consistency of over 95 \% between our approximations and exact analytical results, even though the approximations for small-scale networks (\textit{e.g.}, those with 10 nodes) show noticeable variations and deviations from the exact values, as indicated by a broader distribution and a lower average ratio of the approximation to the exact value.
Such comparisons demonstrate our approximation closely aligns with the mathematically rigorous solution, which yet offers a more explicit form for analytical interpretations and scientific insights in collective cooperation.

\subsection{Pairwise social dilemma within degree-inverse weighted social ties}
The approximated theoretical results enable us to identify a cooperation-enhancing configuration of edge weights under a fixed network connection.
In the pairwise social dilemma, recalling the critical $B^*$ in Eq.~(\ref{bc_star}), we pay more attention to the terms involving the population size $N$ that considerably dominate this mathematical expression due to a typically large value of $N$.
The following component is thus extracted and transformed as
\begin{equation}
	\sum_{u \in \mathcal{N}} \pi_u p_{uu}^{(2)} 
	= \sum_{u,i \in \mathcal{N}} \frac{k_u}{W} \frac{w_{ui}}{k_u} p_{iu} = \sum_{u,i \in \mathcal{N}} \frac{w_{iu}}{W} p_{iu}
	\label{term}
\end{equation}
where $W = \sum_{i \in \mathcal{N}} k_i$ denotes the total sum of weighted degrees on $\mathcal{G}$.
To reduce the threshold $B^*$ required to trigger pairwise cooperation, we seek to amplify the term~(\ref{term}).
This can be mathematically achieved by assigning the higher edge strength $w_{iu}$ to the pair of nodes $i$ and $u$ when the transition probabilities between them (from $i$ to $u$ or vice verse) are large -- typically corresponding to both nodes having relatively few neighbors.
In other words, greater weights are allocated to edges that connect low-degree node pairs.
This weighting process will, in return, reinforce the social ties between such nodes, thereby increasing the possibility of mutual access via the random walk.
Heuristically, we choose to focus on degree-inverse weighted social ties as
\begin{equation}
	w_{ij} \sim (d_{i} \cdot d_j)^{-1}.
\end{equation}
From the theoretical perspective, we argue that the degree-inverse weighted social ties, as opposed to homogeneous counterparts, can more effectively promote cooperation in the pairwise social dilemma.
Meanwhile, we also acknowledge the potential weakness that this organization of edge weights holds a possibility to reduce the weighted degree heterogeneity, $\eta_k$, thereby slightly lowering the threshold of pairwise cooperation.

\begin{figure*}[htbp]
	\centering
	\includegraphics[width=\textwidth]{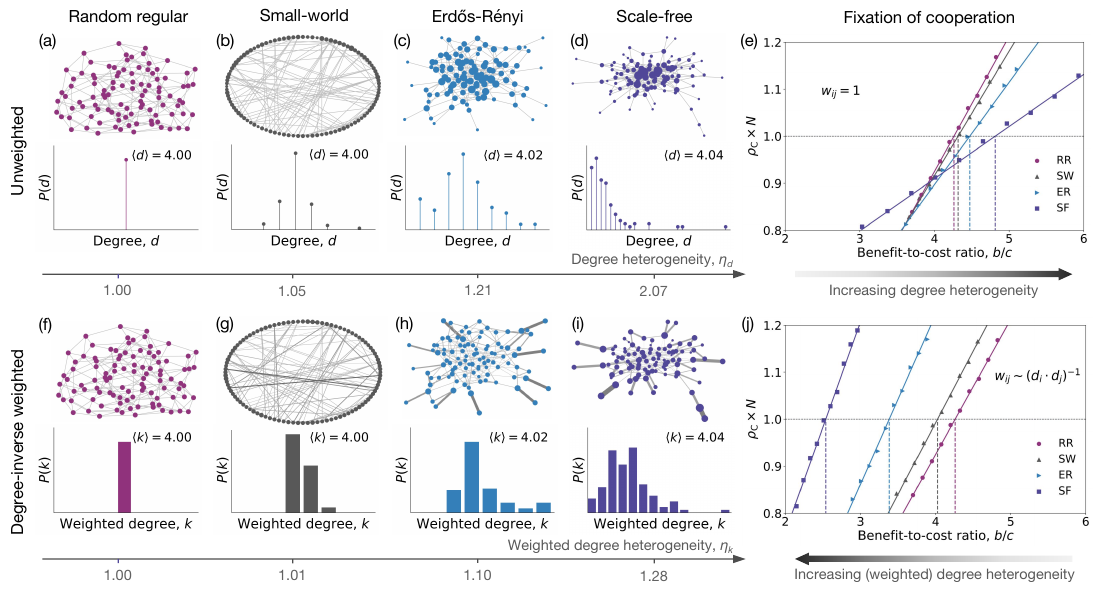}
	\caption{Degree-inverse weighted social ties favor cooperation in the pairwise social dilemma substantially more than unweighted counterparts.
		(a)-(d), Characteristics of random regular (RR, (a)), small-world (SW, (b)), Erd\"os-R\'enyi (ER, (c)) and scale-free (SF, (d)) networks with the size $N = 100$ and the average unweighted degree $\langle d \rangle \approx 4$. 
		These networks show increasing degree heterogeneities, denoted as $\eta_d$, from left to right.
		(e), Scatters of the size-normalized fixation probabilities of cooperation, $\rho_{\mathrm{C}} \times N$, in the pairwise dilemma on networks (a)-(d), versus the benefit-to-cost ratio, $b/c$, of the donation game.
		These are averaged over $2 \times 10^6$ simulation runs with the selection strength $\delta = 0.025$.
		Solid lines represent least-squared fits for each network, with dashed vertical lines indicating the critical benefit-to-cost ratio over which cooperation surpassed the neutral drift (\textit{i.e.}, $\rho_{\text{C}} \times N = 1$).
		Overall, the network with a higher degree heterogeneity is associated with a larger critical benefit-to-cost ratio which is disadvantageous for the propagation of cooperation.
		(f)-\textbf{i}, Characteristics of weighted networks with the same connections between nodes as in (a)-(d) yet the heterogeneous edge weights according to the degree-inverse configuration as $w_{ij} \sim (d_i \cdot d_j)^{-1}$.
		For each network, the average weighted degree $\langle k \rangle$ keeps the same value as the average unweighted degree by rescaling the edge weights.
		Despite the variation of the edge weights, the weighted degree heterogeneities, denoted as $\eta_k$, still exhibit a growing trend from left to right.
		\textbf{j}, Scatters of the size-normalized fixation probabilities of pairwise cooperation on networks (f)-\textbf{i}, versus the benefit-to-cost ratio, achieved by simulations for $2 \times 10^6$ times also with $\delta = 0.025$.
		The fitted lines from the scatters show a reverse relation between the (weighted) degree heterogeneity and the critical benefit-to-cost ratio.}
	\label{fig3}
\end{figure*}

To demonstrate the promotion of such degree-inverse weighted social ties in pairwise cooperation, we examine the fixation process of cooperative behaviors in the donation game on four random networks of size $N=100$ and average unweighted degree $\langle d \rangle \approx 4$.
We start with the networks within homogeneous social ties, as illustrated in Figure~{\ref{fig3}}(a)-(d).
From RR, SW, ER to SF, the heterogeneity of the unweighted degree distribution, $\eta_d$, presents a progressively growing tendency.
Correspondingly, the critical benefit-to-cost ratio, $B^*$, required to support cooperation also rises (Figure~{\ref{fig3}}(e)), yielded by $2 \times 10^6$ parallel simulations for each evolutionary process until the system is fixed.
This reveals that the greater degree heterogeneity tends to undermine the spread of pairwise cooperation, which is consistent with the conventional knowledge of the networked donation game.
However, from Figure~{\ref{fig3}}(f)-(j), degree-inverse weighted social ties can reshape the cooperation ecology with the same set of network connections.
On the one hand, for a fixed network connection (\textit{e.g.}, comparing Figure~{\ref{fig3}}(d) and Figure~{\ref{fig3}}(i)), degree-inverse weighted social ties significantly promote cooperation by lowering the critical benefit-to-cost ratio, despite a noticeable reduction in weighted degree heterogeneity (\textit{e.g.}, from $2.07$ in Figure~{\ref{fig3}}(d) to $1.28$ in Figure~{\ref{fig3}}(i)).
This observation is aligned with our theoretical inference.
Notably, the random regular network shows no change under this weighting rule, as all nodes have identical unweighted degrees.
On the other hand, we find, surprisingly, that networks with higher degree heterogeneity, captured by $\eta_d$ and $\eta_k$, exhibit an even greater propensity to support cooperation under the configuration of degree-inverse weighted social ties, which surpasses the regular network in promotion of cooperative behaviors.
Hence, degree-inverse weighted social ties create a breeding ground for cooperators in the pairwise dilemma across degree-heterogeneous populations.

\subsection{Group social dilemma within degree-inverse weighted social ties}
 In what follows, we turn to the configuration of cooperation-promoting edge weights for the group social dilemma in heterogeneous populations.
Similarly, two terms related to $N$ are extracted from the critical enhancement factor $R^*$~(\ref{r_star}) as
$\sum_{u \in \mathcal{N}} \pi_u p_{uu}^{(0,2)}$ and $\sum_{u \in \mathcal{N}} \pi_u p_{uu}^{(1,2)}$.
Among these, since the first one is akin to the term~(\ref{term}) with the distinction that transition probabilities incorporate self-loops, we have the reason to expect the reinforcing effect of degree-inverse weighted social ties also existed in group cooperation.
The other term, $\sum_{u \in \mathcal{N}} \pi_u p_{uu}^{(1,2)}$, is associated with the three-step recurrence random walk, implying the impact of the connected triplets on the critical threshold of group cooperation.
Here we reveal the importance of how the microscopic degree-related organization shapes the collective cooperation, and thus regard such clustering properties as a significant scientific insight for future research in the networked public goods game.

\begin{figure}[!t]
	\centering
	\includegraphics[width=0.33\textwidth]{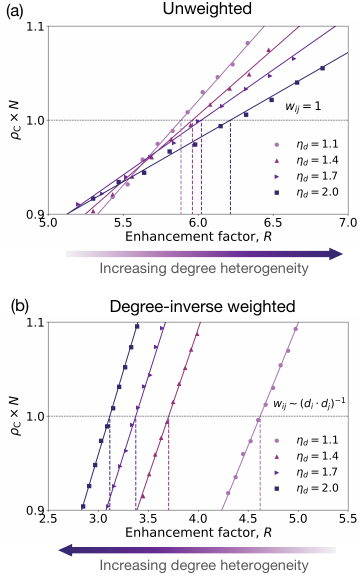}
	\caption{Degree-inverse weighted social ties promote cooperation in the group social dilemma.
		Scatters of the size-normalized fixation probabilities of cooperation in the group dilemma on increasingly degree heterogeneous networks, achieved by $2 \times 10^6$ parallel simulations.
		We apply the network with the size $N = 100$ as well as the average degree $\langle d \rangle = 4$, and adopt the selection strength $\delta = 0.025$ in all simulations.
		The solid and dashed lines indicate the counterparts in the group dilemma, analogous to those in the pairwise dilemma shown in Figure~\ref{fig3}(e) and Figure~\ref{fig3}(j).}
	\label{fig4}
\end{figure}

Subsequently, we investigate the impact of degree-inverse weighted social ties in group cooperation by implementing Monte Carlo simulations to replicate the evolutionary process of cooperation in public goods games across the population.
To eliminate the effect of the clustering coefficient across distinct network types, we focus on a single class of random networks, scale-free networks, with the constant average unweighted degree and varying degree heterogeneities.
As shown in Figure~\ref{fig4}, we discover that degree-inverse weighted social ties also profoundly decrease the threshold of group cooperation for each network, particularly the higher heterogeneous network, and successfully reverse the trend that higher heterogeneous degree distribution is inclined to prohibiting cooperation in the group dilemma.

\subsection{Generality of degree-inverse weighted social ties in promoting cooperation}
\begin{figure*}[!t]
	\centering
	\includegraphics[width=\textwidth]{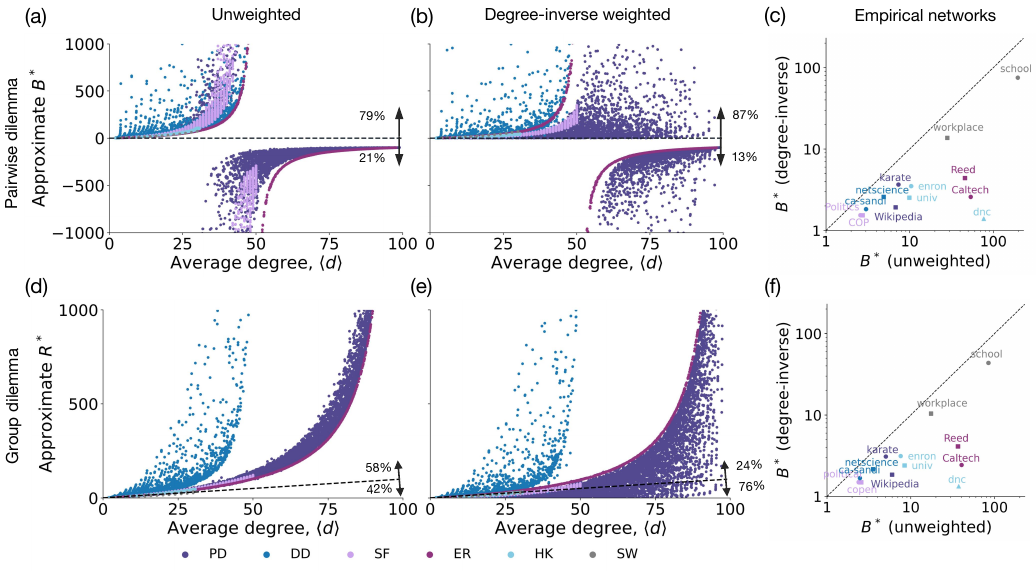}
	\caption{Generality of degree-inverse weighted social ties in promoting cooperation in pairwise and group dilemma on various synthetic and empirical networks.
		There are $30,517$ synthetic networks used in (a), (b), (d) and (e), which are generated using the following algorithms and parameters: 
		$936$ ER with linking probabilities $0.04 \leq p \leq 0.99$; 
		$5,130$ SW with nearest neighbor numbers $4 \leq m \leq 30$ and rewiring probabilities $0.01 \leq p \leq 0.2$; 
		$2,126$ SF using the configuration model with average unweighted degrees $2 \leq d \leq 50$ and power-law exponents $2 \leq \gamma \leq 3$; 
		$5,700$ Holme-Kim (HK)~\cite{holme2002} with added edge numbers $1 \leq m \leq 20$ and edges adding probabilities $0 < p \leq 0.3$; 
		$6,635$ partial duplication (PD) models with initial clique sizes $5 \leq m < 100$, neighbors joining probabilities $0.01 \leq p < 1$ and source nodes joining probabilities $0.01 \leq q < 1$; 
		and $9,990$ duplication divergence (DD) models~\cite{ispolatov2005} with edges retaining probabilities $0.01 \leq p <1$.
		There are $13$ empirical networks~\cite{ryan2015}, categorized by:
		1) Social networks, exemplified by Zachary karate club and Wikipedia who-votes-on-whom, where nodes are members and edges represent social interactions;
		2) Facebook networks, sampled from California Institute of Technology and Reed College, with nodes indicating users and edges capturing friendship ties;
		3) Retweet networks, involving politics and Conference of the Parties (COP), where nodes are twitter users and edges describe retweets linking posters with re-tweeters;
		4) Contact networks, surveyed in the primary school and the workplace, with nodes representing humans and edges describing contact with physical proximity;
		5) Email networks, extracted from Enron, University and Democratic National Committee (DNC), where nodes are users and edges representing email connections;
		6) Collaboration networks, collected from University of California, San Diego, and the research field of network science, with nodes representing authors and edges indicating joint publications between a pair of authors.
		Here DNC email and network-science collaboration networks involve edge weights while the others are unweighted.
		(a)-(b), Scatters of approximate $B^*$ versus average unweighted degree on around $30,000$ synthetic networks with unweighted (a) and degree-inverse weighted (b) edges.
		(c), Comparisons of approximate $B^*$ of pairwise cooperation between homogeneous and degree-inverse edge weights on $13$ empirical networks, categorized into social networks (karate and Wikipedia), Facebook networks (Caltech and Reed), retweet networks (politics and COP), contact networks (school and workplace), email networks (enron, univ and DNC), and collaboration networks (Ca-sandi and netscience).
		All the date dots are shown below the diagonal dash line.
		(d)-(e), Scatters of approximate $R^*$ versus average unweighted degree on the same collection of synthetic networks with homogeneous (d) and degree-inverse (e) edge weights.
		(f), Comparisons of approximate $R^*$ of group cooperation between homogeneous and degree-inverse edge weights on the same set of empirical networks as (c).
	}
	\label{fig5}
\end{figure*}

 To test the generality of degree-inverse weighted social ties in enhancing cooperation, we apply the theoretical predictions of both homogeneous and degree-inverse weighted social ties into nearly $30,000$ synthetic networks of size $100$ (see Figure~{\ref{fig5}}(a), (b), (d) and (e)).
Regarding the pairwise dilemma, the approximate $B^*$ required for cooperation suffers from a general decline with the introduction of degree-inverse weighted social ties (Figure~{\ref{fig5}}(a) -- (b)).
Additionally, the fraction of random networks capable of supporting cooperation (\textit{i.e.}, $B^* > 0$) rises from $79\%$ to $87 \%$ while non-supporting ones drop from $21 \%$ to $13 \%$.
The same trend also manifests in the group dilemma.
As illustrated in Figure~{\ref{fig5}}(d) -- (e), degree-inverse weighted social ties broadly lead to the reduction of the approximate $R^*$ required for group cooperation.
In such networked public goods games, a network structure is considered globally cooperation-inclined if the critical threshold satisfies $0 < R^* < \langle d \rangle + 1$.
This is because $R^* = d + 1$ is the boundary for a $d$-player public goods game, below which defection is dominant over cooperation for each individual.
Correspondingly, the fraction of cooperation-inclined networks increases from $42 \%$ to $76 \%$, while that of defection-inclined networks decreases from $58 \%$ to $24 \%$.
From a wide range of random networks, we demonstrate that degree-inverse weighted social ties can stably reduce the threshold of pairwise and group cooperation, thus improving the prospect of cooperation across the network.

To shed light on cooperation in human society, we investigate $13$ empirical networks~\cite{ryan2015} categorized by social, Facebook, retweet, contact, email, and collaboration networks.
As shown in Figure~{\ref{fig5}}(c) and Figure~{\ref{fig5}}(f), all the data points lie below the diagonal line, suggesting degree-inverse weighted social ties are associated with lower thresholds of both pairwise and group cooperation compared to homogeneous social ties.
Despite this, we find that networks of the same category tend to cluster with one another, due to the similar topological properties from the identical social scenario.
Of all the empirical networks considered, the DNC email network shows the largest shift in the cooperation threshold when transitioning from homogeneous to degree-inverse weighted social ties, since the corresponding scatter is farthest from the diagonal line.
Meanwhile, the Facebook network type, to the greatest extent, is rescued in cooperation by degree-inverse weighted social ties.
These findings strengthen the generality of degree-inverse weighted social ties in enhancing cooperation and provide a valuable insight for network analyses and social governance in the real-world setting.

\subsection{Mechanism of degree-inverse weighted social ties in promoting cooperation}
\begin{figure*}[!t]
	\centering
	\includegraphics[width=\textwidth]{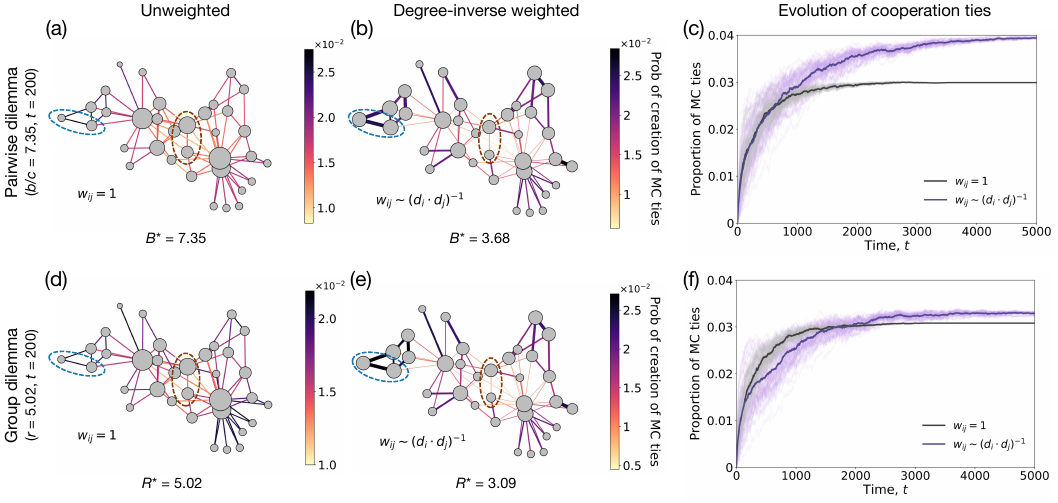}
	\caption{Network snapshots and temporal evolution of cooperation for the unweighted and degree-inverse weighted network in both pairwise and group dilemmas.
		We use the Zachary karate club network, consisting of $34$ nodes and $77$ edges, to compute the approximated $B^*$ for the pairwise dilemma and the approximated $R^*$ for the group dilemma with both unweighted and degree-inverse weighted edges.
		Simulations of the evolution of mutual-cooperation (MC) ties are replicated $10, 000$ times with the benefit-to-cost ratio $b/c=7.35$, the enhancement factor $r = 5.02$, and the selection strength $\delta = 0.025$.
		(a) -- (b), (d) -- (e), visualizations of the average probability of MC tie formation at time $t = 200$ for unweighted and degree-inverse weighted networks in both dilemma types.
		The node size represents the weighted degree, while the edge color reflects the average probability of MC tie creation.
		Notably, edges connecting low-degree nodes (as circled in blue) typically show darker colors (indicating higher probabilities), whereas those linking high-degree nodes (as circled in red) tend to be lighter.
		(c), (f), the average proportion of MC ties over time in the pairwise (c) and group (f) dilemmas, respectively, comparing unweighted (gray) and degree-inverse weighted (purple) networks.}
	\label{fig6}
\end{figure*}

 To fundamentally unveil the mechanism of degree-inverse weighted social ties in promoting cooperation, we explore the formation and evolution of mutual-cooperation (MC) ties exemplified by the Zachary karate club network with $34$ nodes and $77$ edges (Figure \ref{fig6}).
In the pairwise dilemma, the theoretical threshold $B^*$ decreases from $7.35$ to $3.68$ after the introduction of degree-inverse weighted social ties.
We firstly visualize the probability of creation of MC ties across the edges of the population at time $t = 200$, achieved by  $10, 000$ simulations.
As illustrated in Figure~\ref{fig6}(a), it is apparent that the edge connecting both nodes with lower unweighted degrees is prone to the emergence of MC ties (as indicated by the blue circle) while the connected hubs seem difficult to realize mutual cooperation (as indicated by the red circle).
Degree-inverse weighted social ties further amplify the prospect of the MC tie between low-degree nodes, thus driving the population towards the cooperation-favorable state (Figure~\ref{fig6}(b)).
From the global viewpoint, the average proportion of MC ties is higher in degree-inverse weighted social ties, naturally more conducive to cooperation, by comparison with homogeneous social ties.
These are also confirmed in the group dilemma.
Overall, we demonstrate the facilitation of the formation of cooperator ties from the perspective of evolutionary mechanisms in both pairwise and group dilemmas.

\section{Discussion}
In this work, we proposed a general framework for determining the conditions for cooperation in both pairwise and group social dilemmas, which can be applied to any weighted network.
By integrating the quenched mean-field approximation with the coalescing theory, we derived interpretable critical thresholds, over which cooperation is favored to evolve in the structured population.
Our theoretical predictions were validated through extensive Monte Carlo simulations across a wide range of synthetic network models with various weighting schemes, and also by comparison with the exact solution derived from \cite{allen2017} on varying-scale networks.

Such approximations provide novel theoretical insights into the fundamental principles underlying the emergence and evolution of cooperation in heterogeneous populations.
Although previous works have developed the mathematically rigorous solution, the associated computational complexity poses significant challenges for analyzing large-scale networks.
Moreover, the complicated solution often inhibits the quantitative understanding of how specific network properties shape evolutionary dynamics.
In contrast, our analytical framework enables an explicit interpretation of network effects through the lens of random walks, providing a unifying approach that applies to both pairwise and group interaction scenarios. 
By balancing accuracy with interpretability, our method facilitates a deeper understanding of the rationale behind collective cooperation and contributes an effective tool for the quantitative analysis of evolutionary games on networks.

Furthermore, theoretical analyses motivate us to consider degree-inverse weighted social ties can reduce the analytical thresholds of both pairwise and group cooperation, thus enhancing collective cooperation in the structured populations.
We further confirmed the generality of this rule by testing it on various synthetic and empirical networks. 
Ultimately, we offered a mechanistic perspective to reveal how degree-inverse weighted social ties foster cooperation by facilitating the formation and stability of cooperative ties.
The rule of configuring social ties underscores the pivotal role of network topologies in steering population dynamics toward collective benefit.

Another noteworthy insight from our study is that degree heterogeneity does not necessarily undermine the proliferation of cooperation, particularly on degree-inverse weighted networks in which stronger ties are concentrated between less-connected nodes while weaker ties are assigned between hubs.
This observation offers a meaningful refinement to the conventional view in evolutionary game theory, which generally holds that degree heterogeneity suppresses cooperative behavior.
Beyond the degree-inverse weighting scheme, prior research has shown that heterogeneous networks indeed outperform homogeneous ones in enhancing cooperation when individual update rates vary inversely with their degrees~\cite{meng2024}.
Moreover, Meng et al.~\cite{meng2024arxiv} proposed a simple rule of thumb for the emergence of cooperation: hubs should be temporally deprioritized in interactions relative to peripheral nodes from a time-varying perspective.
Our scientific finding adds a valuable dimension to the growing body of cooperation-favoring mechanisms and carries meaningful implications for macro-level interventions, including institutional design, policy-making, and the governance of collective behavior in complex social systems~\cite{Farahmand2010, Mulder2018}.

Despite the theoretical findings and implications of our study, several limitations should be discussed.
First, our theoretical framework is developed under the assumption of weak selection, suggesting a small influence of game interactions on dynamical evolution.
Extending the analysis to strong selection regimes, however, remains challenging due to the complex interdependence of strategic states, and warrants further mathematical development.
Second, our analysis is confined to simple network structures. 
In contrast, real-world systems often exhibit multi-layer, higher-order~\cite{Preen2019}, or time-varying features~\cite{Lones2014, dinh2015}, which offer a more faithful representation of social and biological interactions.
Investigating how heterogeneous social ties reshape cooperative dynamics in such more intricate networked settings is also a promising topic worthy of academic attention.
Thirdly, we solely treat tie strengths as static attributes of the network. 
Yet, in practice, social engagement between individuals can evolve over time, or even co-evolve with behavioral strategies~\cite{wei2024, Aghbolagh2023}, which presents a promising direction for further exploration.
Lastly, beyond the scope of one-shot games considered in this study, the dynamics of iterated games in structured populations also present an important and intriguing direction for future research~\cite{chen2023, chiong2012, chong2007}.
Understanding how weighted edges shape evolutionary success in repeated interactions remains an essential question worthy of deep consideration.
Taken together, promising future research could be developed alongside these avenues to enhance our understanding of the emergence and persistence of cooperation in heterogeneous populations.

	\vspace{5mm}
	\textbf{\large Corresponding author}: Correspondence to Dini Wang (Dini.Wang@dartmouth.edu), Peng Yi (yipeng@tongji.edu.cn), or Feng Fu (Feng.Fu@dartmouth.edu).
	
	\vspace{5mm}
	\textbf{\large Competing interests}: The authors declare no competing interests.

\clearpage

\clearpage

\clearpage

\clearpage

\clearpage

\clearpage

\clearpage

\begingroup
\bibliographystyle{naturemag} 
\bibliography{citation_arxiv}
\endgroup

\end{document}